\documentclass[aps,letterpaper,twocolumn,pre,floatfix]{revtex4-1}
\usepackage{graphicx}
\usepackage{bm}
\usepackage{hyperref}
\usepackage{amssymb}
\usepackage{amsmath,mathtools}
\usepackage{epsf}
\usepackage{esint}
\usepackage[caption=false]{subfig}
\usepackage{epstopdf}
\usepackage{amsfonts}
\usepackage{color}
\DeclareMathOperator\tr{tr}
\begin{document}
\title{Geometric phaselike effects in  a quantum heat engine}
\author{Sajal Kumar Giri}
\author{Himangshu Prabal Goswami}
\email{hpgoswami@pks.mpg.de}
\affiliation{Finite Systems Division, Max-Planck-Institute for the Physics of Complex Systems, Dresden, Germany}
\date{\today}

\begin{abstract} 
 By periodically
 driving the temperature of reservoirs in a quantum heat engine, geometric or Pancharatnam-Berry phase-like (PBp)  effects in the
 thermodynamics can be observed. The PBp can be identified from  a generating function (GF) method within an 
 adiabatic quantum Markovian master equation formalism. The GF is shown not to lead to a standard open quantum system's fluctuation theorem in presence of phase-different
 modulations 
 with an inapplicability in the use of large deviation theory.
 Effect of quantum coherences in optimizing the flux is nullified due to PBp contributions.
 The linear coefficient, $1/2$, which is universal
 in the expansion of the efficiency at maximum power in terms of Carnot efficiency no longer holds true in presence of PBp effects.  
\end{abstract}
\maketitle

\section{ Introduction}
  The Pancharatnam-Berry phase or geometric phase (GP) \cite{berry1984quantal,pancharatnam1956generalized}
in quantum systems  is
a natural emergence when the system is subjected to local periodic modulations of its structural parameters. In open quantum systems too,
 parametrization of the system's parameters such as tunneling coefficients \cite{kaestner2015non}, applied potential \cite{switkes} 
as well as temperatures \cite{watanabe2014non} have 
been known to affect open quantum systems' observables  \cite{pluecker2017gauge}. 
Periodically driven temperatures in an anharmonic junction
have been shown to lead to fractional quantization 
     of the heat flux \cite{ren2010berry}. Such non-trivial thermodynamics arising due to periodic modulation of temperatures
     have also been studied formally in classical Brownian heat engines where effects on
     entropy and Onsager's coefficients 
    were investigated \cite{PhysRevX.5.031019} as well as in a nonequilibrium spin boson model\cite{PhysRevA.95.023610}.
    Recently, 
    quantum mechanical versions of heat engines where the working medium is discretized \cite{scovil1959three} have 
    resurfaced\cite{quan2007QHE,kosloff2013quantum,gelbwaser2017thermodynamic}. 
The  interest in quantum heat engines (QHE)  has a twofold directive, firstly
 to answer the basic question, 
 when the heat engine is really quantum \cite{PhysRevX.5.031044} and secondly to understand the thermodynamics 
when the quantum effects like coherences and entanglement are predominantly active in the engine \cite{scully2003extracting,4qheEnta,gelbwaser2015strongly}. 
Several intriguing works exist, such as extracting work from a single bath, increase in power aided by noise-induced quantum 
coherence \cite{scully2003extracting,ScullyPNAS13092011}, surpassing Carnot efficiency using 
squeezed versions of thermal baths 
\cite{rossnagel2014nanoscale} as well as experimental realization of a Paul trapped single $Ca^+$ ion heat engine \cite{rossnagel2016single}. QHEs have also been experimentally 
realized in  laser cooled Rb atoms\cite{PhysRevLett.119.050602} using principles based on electromagnetically induced transparency\cite{EITQHEharris}.
Despite the progress, the issue of GP in QHEs have not yet been addressed, both within and 
    beyond the linear response regime of finite-time thermodynamics. 
In this work we focus on a QHE where one can realize 
Pancharatnam-Berry phase-like (PBp) contributions through periodically modulating the temperature of the thermal baths.

A heat engine's performance is analyzed by evaluating its efficiency and power. 
Efficiency, in the finite power regime, needs to be calculated by maximizing the power with respect to some parameter of the system. This is known as the efficiency 
 at maximum power ($\eta_*$), theoretically given by, $\eta_*=1-\sqrt{1-\eta_c}$\cite{Curzon-Ahlborn, Novikov1957}, where $\eta_c=1-T_c/T_h$ is the Carnot efficiency and
 $T_c(T_h)$ represents cold (hot) thermal bath's temperature.
Close to equilibrium in the endo-reversible regime, $\eta_*=\eta_c/2+\eta_c^2/8$, with the coefficient
$1/2$ being universal \cite{seifert2012stochastic,PhysRevLett.102.130602,BroeckEMP}. 
  Further it is also known that 
  the efficiency at maximum power is bounded above and below by
  $\eta_c/2\le\eta_*\le \eta_c/(2-\eta_c)$ \cite{izumida2012efficiency,PhysRevE.87.012133}. 
  In this work, we show that  this strong standing  linear expansion
 coefficient, $1/2$, doesn't hold due to the emergence of PBp effects and one can 
  go both above and below the upper bound on $\eta_*$ by having phase different driving protocols. 
  Further, a QHE works by absorbing heat from the reservoirs in the form of quanta 
which  mimics boson exchange, making them ideal to study heat transfer as a quantum transport phenomena from which the role of random fluctuations are studied. 
Random fluctuations affect the transport properties and are usually studied via a full counting statistics (FCS) 
\cite{Levitov,nazarov} method that involves calculation of moments and cumulants of $P(q,t)$, the probability distribution function (PDF) for the
  number ($q$) of particles exchanged 
  between system and reservoir in a measurement time $t$.
FCS led to the steady state fluctuation theorems (FT) \cite{
seifert2012stochastic,campisi2011colloquium,uhrmp,SaarQHEPhysRevA.86.043843,UHeplQHE,campisi2015nonequilibrium}, the cornerstones of
quantum thermodynamics \cite{kosloff2013QTherm}. 
         Recently, it was reported that during  transport across
  quantum junctions \cite{ren2010berry,hpg4}, the standard mathematical 
  form of the FT is broken due to the emergence of PBp and hence attempts have been made to establish geometric 
   FTs \cite{PhysRevE.96.022118} in spin-boson systems. We observe the same violation of the FT in our QHE. Further,  the long time PDF 
  is evaluated by invoking the use of large deviation theory \cite{touchette2009large,varadhan1984large} and steadystate (SS) FTs are
  derived from large deviation results \cite{uhrmp,dhar2,lebowitz1999gallavotti}. We find that the large deviation theory cannot be used
   to determine the PDF in presence of PBp contributions.

  In this work we focus on the effects of PBp contributions on the
    thermodynamics of a coherent QHE beyond the linear response regime, where  the temperature of the two thermal baths are periodically modulated in time.
       The technique used in this paper is a standard 
       procedure based on a quantum Markovian master equation (QMME) formalism (weak coupling between system and bath) 
       combined with a generating function method popularly used in FCS.  The paper is organized as follows. In Sec. (\ref{qhe}), we introduce the model QHE,
       derive a QMME and discuss the PBp curvature. In Sec. ( \ref {s3}), we look at the thermodynamics of the QHE 
       focusing on the geometric contributions to flux and efficiency at maximum power. In Sec. (\ref{s4}), we analyze the PBp effects on the SSFT and large deviation theory
       which is followed by conclusions and an appendix.

\section{Engine Specifications}
\label{qhe}
\begin{figure}[!tbp]
\centering
\includegraphics[width=7.5cm]{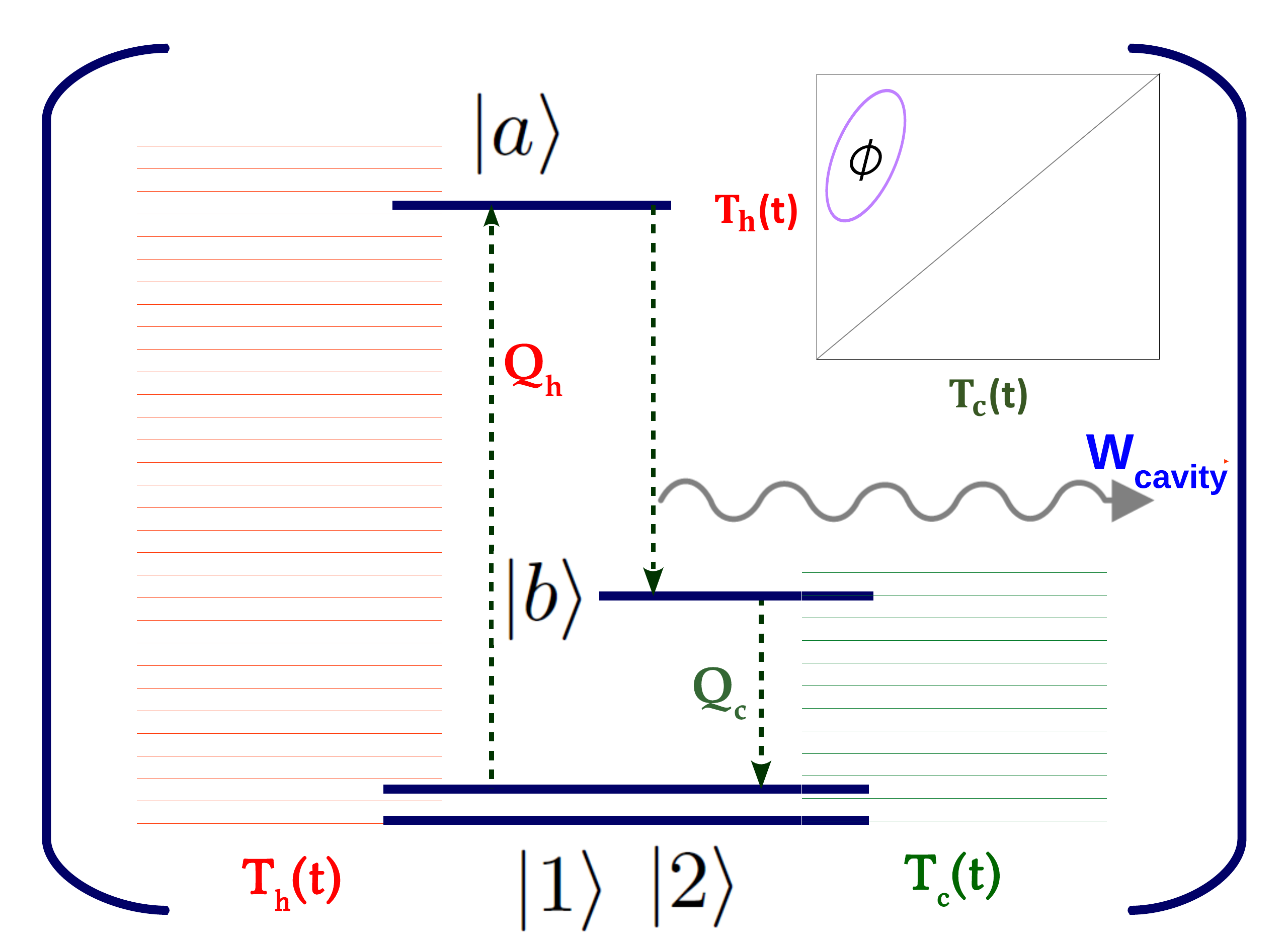}
\caption{ A four level quantum heat
engine. Degenerate levels $|1\rangle$ and $|2\rangle$ are coupled
to two excited levels $|a\rangle$ and $|b\rangle$ through the  two driven thermal 
baths  at temperatures $T_h$(t) and $T_c$(t). Levels
$|a\rangle$ and $|b\rangle$ are coupled to a cavity with a single mode
with a frequency $\nu_l$. Emission of photons into this mode is the work
done by the QHE. The PBp is quantified by the area traced in the parameter space of the two temperatures with nonzero phase difference, $\phi (\phi\ne n\pi, n\in$ integers).
}
\label{c7-schematic}
\end{figure}
The QHE model that we study has  also been studied in Refs. \citep{SaarQHEPhysRevA.86.043843,UHeplQHE,goswami2013thermodynamics,ScullyPNAS13092011}
 and is shown in Fig. (\ref{c7-schematic}).
The model consists of two thermal baths at temperatures $T_h(t)$ and $T_c(t)$. The two temperatures are periodically driven in time such that, $T_h(t)>T_c(t).$
There are two degenerate states, $|1\rangle$ and $|2\rangle$,
with forbidden transition and are coupled to the two time varying thermal baths. The higher energy state $|a\rangle (|b\rangle)$ is coupled to hot (cold) bath. 
The two higher states, with an allowed transition, $|a\rangle$ and $|b\rangle$ are also coupled to a unimodal cavity. 
The total Hamiltonian can be written as $\hat{H}_T= \hat{H}_o+\hat{V}$, where
\begin{equation}
\hat{H}_o=\displaystyle\sum_{\nu=1,2,a,b}
E_\nu\hat{B}_{\nu\nu}+\displaystyle\sum_{k\in h,c}\epsilon_k^{}\hat{a}_k^\dag\hat{a}
_k+\epsilon_l\hat{a}_l^\dag\hat{a}_l.
\end{equation}
Here, $\nu=1,2,a,b$, $E_\nu$ is the energy of the $\nu$th level and $\hat{B}_{\nu \nu^\prime }=|\nu \rangle\langle \nu^\prime |$ is
the  operator that causes excitation
between the states $|\nu \rangle$ and $|\nu^\prime \rangle$. $\epsilon_k$ represents the energy of $k$th mode of
the hot(h) and cold (c) thermal reservoirs and $\epsilon_l$ represents the energy of the unimodal cavity. 
We model the thermal baths as harmonic modes, where
$\hat{a}^\dag (\hat{a})$ stand for the bosonic creation (annihilation) operators.
The interacting Hamiltonian is $\hat V=\hat V_{sb}^{}+\hat V_{sc}$, with the coupling between
the (working) system and thermal baths being, 
\begin{equation}
\hat{V}_{sb}^{}=\displaystyle\sum_{k\in h,c}\sum_{i=1,2}\sum_{x=a,b}g_{ik}\hat{a}_k\hat{B}_{ix}^\dag+ h.c,
\end{equation}
where $g_{ik}$ is the strength of coupling between the $i$th level and $k$th mode. The coupling between the system and cavity mode is
\begin{equation}
\hat{V}_{sc}^{}=g(\hat{a}_l^\dag\hat{B}_{ba}+\hat{B}_{ba}^\dag\hat{a}_l).
\end{equation}
The strength of coupling between the system and the unimodal cavity
is denoted by $g$. 
The system absorbs heat in the form of quanta from the hot bath and releases it in the cold bath after undergoing a radiative transition from $|a\rangle\rightarrow|b\rangle$.
Emission of coherent photons into the cavity as a result of this transition is the work done.

\subsection{Quantum Markovian Master Equation}
\label{s2a}
Within an adiabatic approximation, a weakly coupled quantum system's dynamics is governed by the quantum Liouville equation,
$
 |\dot\rho(t)\rangle=\breve{\mathcal L}(t)|\rho(t)\rangle,
$
where $|\dot\rho(t)\rangle$ is the time rate of change of the reduced density vector for the
system\cite{njp}.
$\breve{\mathcal L}(t)$ is the Liouvillian superoperator containing the time dependent driving and is responsible for system's evolution. 
Using standard perturbation theory within the Born-Markov approximation, 
the reduced density vector for the QHE is $|\rho\rangle=\{\rho_{11},\rho_{22},\rho_{aa},\rho_{bb},\Re (\rho_{12})\}$, where $\rho_{ii},i=1,2,a,b$ represent populations 
 of the system's many body states and $\Re(\rho_{12})$ is the thermally (noise) induced coherence between the degenerate states $|1\rangle$ and $|2\rangle$. 
   We assume that  the bath relaxation is much faster as compared
    to the driving time. 
    In order to quantify the flux into the cavity mode, we 
    focus on the statistics of the number of photons exchanged between the system and the  cavity.
    Within a measurement window, $t$,
the PDF corresponding to $q$ net photons in the cavity is $P(q,t)$. The statistics of $q$ is quantified by the 
 moment generating function, defined as $G(\lambda,t)=\sum_qe^{\lambda q}P(q,t)$. From a general framework of FCS, it can be shown that 
 the equation of motion for $G(\lambda,t)$ is
$
   \dot G(\lambda,t)=\langle \breve{{\boldsymbol 1}}| \breve{\mathcal L}(\lambda,t)|\rho(\lambda,t)\rangle
$\cite{dhar2,Levitov,uhrmp,uhFTfcs}.
 $\lambda$ is the auxiliary field that counts the number of photons exchanged between the system and the cavity. 
  The transformed characteristic counting
  Liouvillian is given by (appendix), $\breve{\mathcal L}(\lambda,t)=$
  \begin{equation}
  \label{Liouvillian-lambda}
r\begin{pmatrix}
n(t)&0&\tilde{n}_h(t)&\tilde{n}_c(t)&y(t)\\
0&n(t)&\tilde{n}_h(t)&\tilde{n}_c(t)&y(t)\\
n_h(t)&n_h(t)&\frac{-g^2\tilde{n_l}-2r\tilde{n}_h^{}(t)}{r}&\frac{g^2n_l e^{-\lambda}}{r}&2p_hn_h(t)\\
n_c(t)&n_c(t)&\frac{g^2\tilde{n_l}e^{\lambda}}{r}&\frac{-g^2n_l-2r\tilde{n}_c(t)}{r}&2p_cn_c(t)\\
\frac{y(t)}{2}&\frac{y(t)}{2}&p_h\tilde{n}_h(t)&p_c\tilde{n}_c(t)&n(t)-\tau
\end{pmatrix}.\\[2mm]
\end{equation}
We have  denoted the time and counting-field dependent density vector as
$|\rho(\lambda,t)\rangle$ and $\langle\breve{{\boldsymbol 1}}|=\{1,1,1,1,0\}$. All
couplings between the QHE and thermal baths are 
assumed to be equal and denoted by $r$. The adiabatic limit is valid for $rt'\gg 1$, where
$t'$ is the time scale of the external drivings.  Also, $\breve {\mathcal L}(\lambda=0,t)=\breve{\mathcal L}(t).$
Here, $n(t)=-n_c(t)-n_h(t)$, $y(t)=-n_c(t)p_c-n_h(t)p_h$, with 
  $n_x(t),x\in h,c$, $\tilde{n}_x(t)=n_x(t)+1$. These are given by,
$
 n_c(t)=(\exp\{(E_b-E_1)/k_BT_c(t)\}-1)^{-1}, n_h(t)=(\exp\{(E_a-E_1)/k_BT_h(t)\}-1)^{-1}
$.
 The occupation of the cavity mode is
$
 n_l=1/(\exp[(E_a-E_b)/k_BT_l]-1), \tilde n_l=1+n_l
$, $T_l$ being a fictitious temperature of the cavity\cite{goswami2013thermodynamics}.
The dimensionless parameters, $p_h$ and $p_c$ represent quantum coherence control parameters associated with the hot and cold 
baths respectively\cite{goswami2013thermodynamics,SaarQHEPhysRevA.86.043843,ScullyPNAS13092011}. These two parameters are a measure of interference 
between the transitions involving the ground states and the two excited states.
$\tau$ is a dimensionless dephasing rate introduced phenomenologically \cite{ScullyPNAS13092011,SaarQHEPhysRevA.86.043843} so as to take care of environmental dephasing effects.

\subsection{Pancharatnam-Berry Curvature}
\label{s2b}
In the long time limit, PBp contributions can be realized in
the scaled cumulant generating function given by $S(\lambda)=\lim_{t\rightarrow\infty}(1/t)\ln G(\lambda,t)$. 
 $S(\lambda,t)$ is
 additively separable into two parts\cite{sin} (appendix)
  $S(\lambda,t)=S_d(\lambda,t)+S_g(\lambda,t)$,
 \begin{eqnarray}
 \label{s-dyn}
S_d(\lambda)&=&\frac{1}{t_p}\displaystyle\int_0^{t_p}dt'\zeta_o(\lambda,t'),\\
S_g(\lambda)&=&-\frac{1}{t_p}\int_0^{t_p}\langle L_o({\lambda,t})
 |\dot R_o({\lambda,t})\rangle dt.
 \label{s-geo1}
 \end{eqnarray}
 Here, $S_{d(g)}(\lambda)$ denotes the dynamic(geometric) cumulant generating function. $|R_o(\lambda,t)\rangle $ and $\langle L_o(\lambda,t)|$ denote the instantaneous
 right and left eigenvectors of $\breve{\cal L}(\lambda,t)$ corresponding to the instantaneous long-time dominating
 eigenvalue, $\zeta_o(\lambda,t)$. Converting the line integral to a contour integral in the parameter space of $T_c(t)$ and $T_h(t)$ over a contour ${\cal C}$ and 
 assuming 
 the contour to be closed (a
 fixed time period) and piecewise smooth, we 
can use Stokes' theorem and 
rewrite the contour integral as a surface integral
 enclosed by the closed loop $\mathcal S$,
 \begin{align}
\label{s-geo}
 S_g(\lambda)&=-\frac{1}{t_p}\oiint_{\mathcal S}\nabla\times\langle L_o({\lambda, \bf T})
 |\partial_{{\bf T}}|R_o({\lambda,
 \bf T})\rangle. d{\mathcal S}.\nonumber\\
 \end{align}
 Here, vector ${\bf T}$ contains system parameters modulated by the external driving in the contour.
Eq. (\ref{s-geo})  has a
geometric interpretation since it is quantified by the surface's area and is both re-parametrization as well as gauge invariant\cite{sin, ren2010berry,hpg4}.
The geometric interpretation of Eq. (\ref{s-geo}) is analogous to the original GP interpretation in 
 isolated quantum dynamics \cite{berry1984quantal,mukunda1993quantum,aharonov1987phase}, albeit not being a phase.
 We simply refer to it as the geometric phase-like or Pancharatnam-Berry phase-like  contribution.
The measurement time,
 $t=n t_p$, where $n$ is the number of cycles and
  $t_p$ is the time-period of the driving such that $r t_p\gg 1$.
$\langle L_o({\lambda, \bf T})
 |\partial_{{\bf T}}|R_o({\lambda,
 \bf T})\rangle$ is equivalent to 
 the geometric potential in the parametric  space of $T_c(t)$ and $T_h(t)$ whose curl gives the curvature. It vanishes when there is no phase-difference between 
 the driving protocols (also for $\phi$ as integer multiples
  of $\pi$) since there is no area traced 
  in the parameter space.

\section{Thermodynamics}
\label{s3}
      \begin{figure}
\centering
\includegraphics[width=8.5cm]{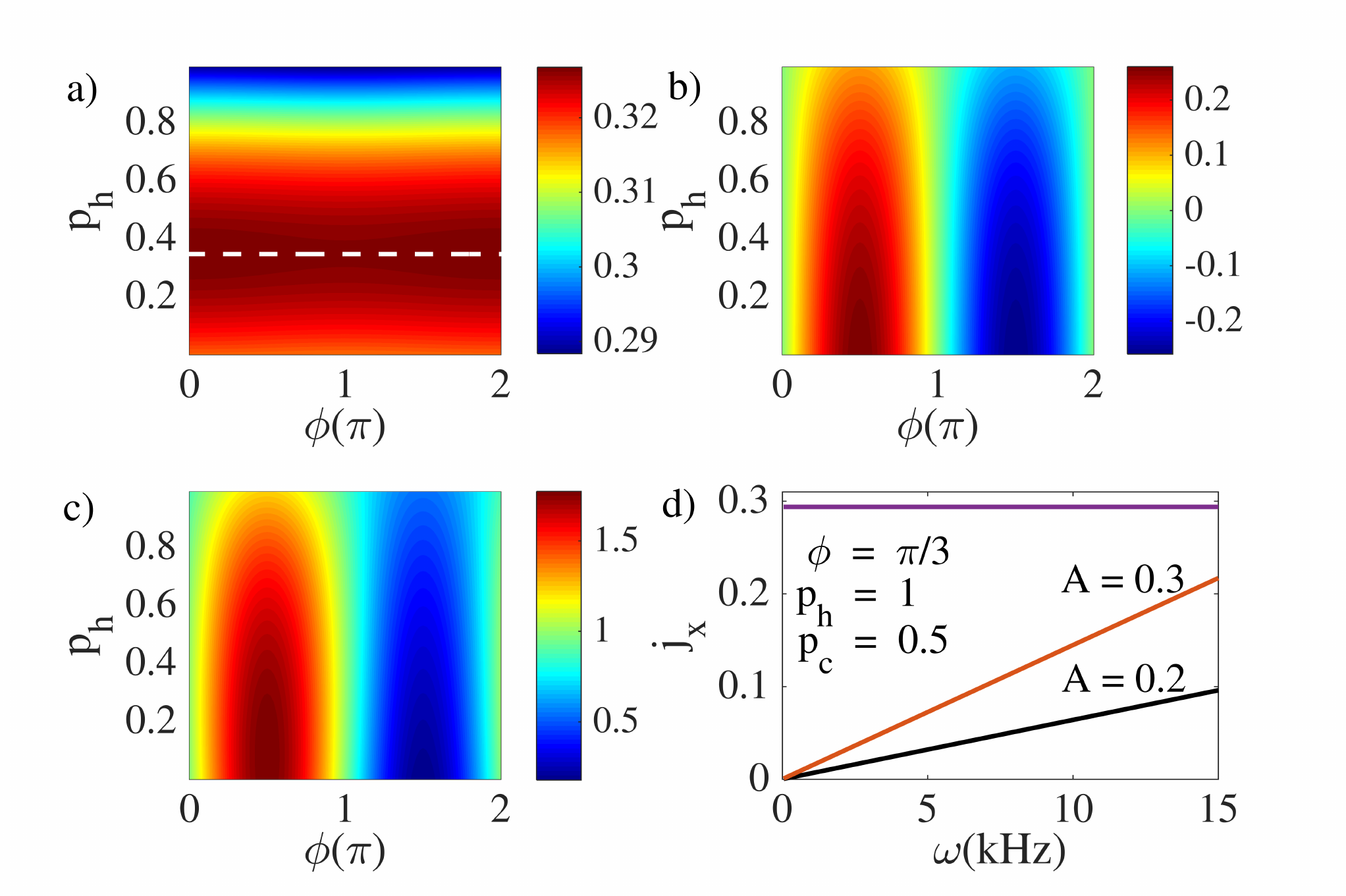}
\caption{  a) Dynamic flux contour. Dashed line indicates that $p_{h,d}^*$ is constant along $\phi$. 
b) The geometric flux contour. 
The optimal values occur at zero coherences. c) $j/j_o$ contour (dimensionless). Due to higher geometric contribution, the optimal flux values
occur at zero coherence value. $E_1=E_2=0.1,E_b=0.4,E_a=1.5,r=0.7,g=10,\tau=0.5,p_c=0.4,\omega=1.2,T_h=4,T_c=2,T_l=1,A=0.2.$
d) Linear (no) dependence of geometric (dynamic) flux on driving frequency.
 Energies ($E_1,E_2,E_b,E_a$), temperatures ($T_h,T_c,T_l$) are in atomic units ($\hbar=1,k_B=1$), rates $g^2,r$ have the 
 units of time inverse and the driving frequency, $\omega$ is in $kHz$ throughout all figures.
}
\label{jdyn}
\end{figure}
\begin{figure}
\centering
\includegraphics[width=8.5cm]{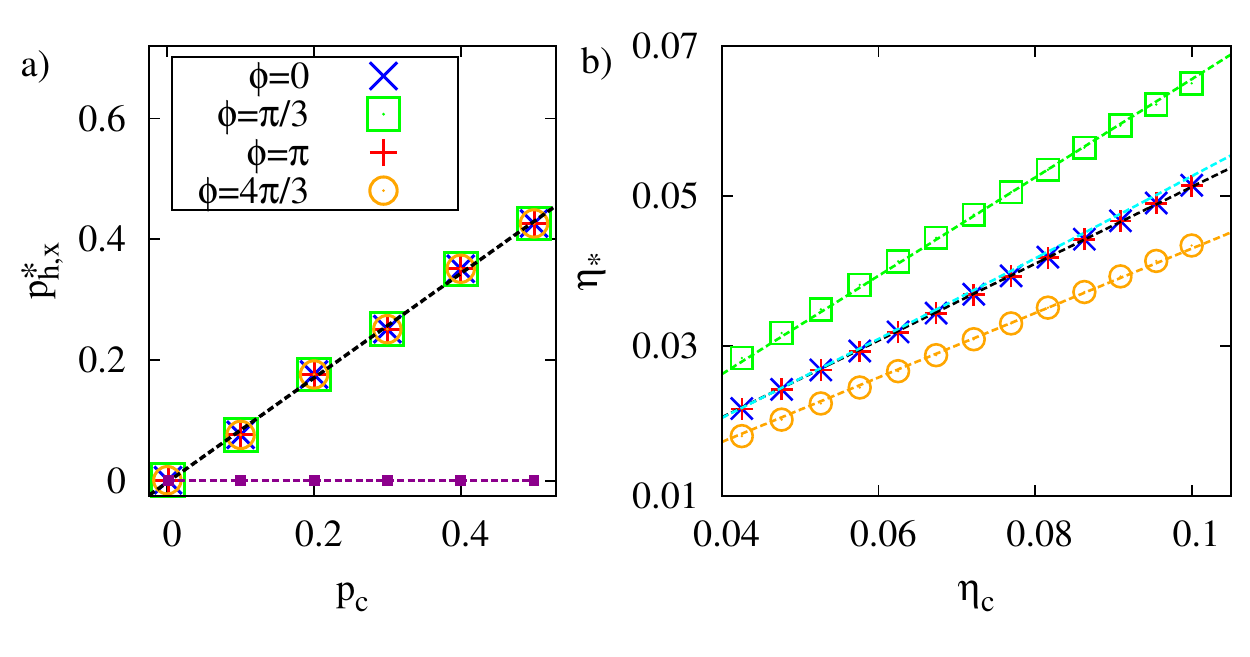}
\caption{a) Plot showing the linear (no) dependence of $p_{h,d}^* (p_{h,g}^*)$ on $p_c$ and both 
 are independent of the phase difference ($p_{h,g}^*$ is the dashed magenta line).
b) Near equilibrium EMP as a function of Carnot efficiency. Only for $\phi=0,\pi$, the slope is 1/2.
For other phase difference, linear slopes greater ($\phi=\pi/3$, slope=0.65) and lower ($\phi=4\pi/3$, slope=0.42)
than 1/2 are observed. Dashed black (cyan) line is the lower (upper) bounds on EMP, $T_c=0.9$, $T_l=2$, $p_c=0.1$, $p_h=1$, $g=50$, $\omega=3,$ $A=0.02, r=0.7,\tau=0.5$.
}
\label{ph-linear}
\end{figure}

We now focus on the flux, work and efficiency of the QHE. 
The flux is defined as the rate of change of the number of photons exchanged between the system and the unimodal cavity. It can be obtained 
 using, $j=\partial_\lambda S(\lambda)|_{\lambda=0}$\cite{uhrmp}. Since $S(\lambda)$ is separable into a geometric and dynamic part, 
  the flux is a sum of   dynamic ($j_d$) and geometric ($j_g$) contributions. 
  For sinusoidal drivings $T_h(t,\phi)=T_h(1-A\sin(\omega t+\phi))$ and $T_c(t)=T_c(1-A\sin(\omega t))$, with the amplitude $A<1$, $\omega$ as the driving frequency and 
   $\phi$ as the phase difference between the two driving protocols, we
   numerically  evaluated the flux which is shown in Fig.(\ref{jdyn}). 
 Similar to a non-driven engine\cite{goswami2013thermodynamics}, the dynamic 
 flux can be maximized with respect to the coherence parameter $p_h$ (dashed line in Fig.(\ref{jdyn}a) for a fixed $p_c=0.4$).
   Note that, this coherence-maximized value of the dynamic flux (at constant $p_c$) as a function of phase difference is always at a constant $p_h$ value, i.e at $p_h=p_{h,d}^*$. 
  In Fig. (\ref{jdyn}a), at $p_h=p_{h,d}^*(=0.35)$, the coherence-maximized value of the flux decreases and reaches a minimum value at $\phi=\pi$ 
  (where the two drivings are completely out of phase and the thermodynamic force is minimum). Infact for a fixed value of $p_h$, the flux always decreases as a function of $\phi$, 
  being minimum at $\phi=\pi$ before increasing again. This trend is robust for all values of $p_c$ and $p_h$.
  Also,
    $p_{h,d}^*$ for a fixed set of other parameters
     linearly depends on $p_c$ and this linearity is also independent of the phase difference as shown 
    in Fig.(\ref{ph-linear}a).  This linear dependence of $p_{h,d}^*$ on $p_c$ was also observed in the nondriven heat engine \cite{goswami2013thermodynamics}.
      


  The interesting quantity, however, is the geometric flux. 
  It doesn't typically follow the same dependence on the phase difference as the dynamic flux does. As can be seen from the contour Fig. (\ref{jdyn}b), 
   the geometric flux, oscillates as a function of the phase difference. For any arbitrary $p_h$, we see that it increases from zero ($\phi=0$ 
   ) and reaches a maxima at $\phi=\pi/2$ before decreasing to become zero again at $\phi=\pi$. After that it is negative, reaches a minima at $\phi=3\pi/2$ and
   then increases to zero at $\phi=2\pi$. 
   The negative component of the geometric flux has been utilized in creating pumping in quantum transport \cite{ren2010berry,hayakawa}.
   In our case, we further observe that the value of $p_h$ that optimizes  the geometric flux, $p_{h,g}^*$, 
   is always at $p_{h,g}^*=0~\forall ~p_c,\phi$ as
    shown in Fig. (\ref{ph-linear}a, dashed magenta line). 
      This  indicates 
     that coherences are exclusive to optimizing the  dynamic flux alone.
  Since the geometric flux increases linearly with $\omega$, Fig.(\ref{jdyn}d) and the dynamic flux 
      doesn't depend on $\omega$, the contribution to total flux is controllable through $\phi$ and $\omega$.
           In the limit of high $\omega$ and nonzero $\phi (\phi\ne n\pi,n\in$ integers), $j_g$  contributes dominantly to the total flux and 
     the global maxima (minima) of the total flux occurs when there is no coherence. This is shown in Fig. (\ref{jdyn}c), where the global maxima (minima)
     of $j/j_o$ ($j_o$ is the flux  when $\phi=0, p_c=p_h=0$ or $ p_c=p_h$)
      is at $p_h=0,\phi=\pi/2(3\pi/2)$. In this limit, the optimum value of the total flux occurs in absence 
       of quantum coherences. 
     Quantum coherences are known to optimize the flux and power of QHEs beyond classical values\cite{ScullyPNAS13092011} in quantum heat engines.
      However, by increasing the geometric flux, one can nullify the effect of quantum coherences in optimizing the total flux by increasing the frequency of driving.
      Note that the upper limit of $\omega$ is in the $THz$ regime beyond which the adiabatic approximation will fail\cite{ren2010berry}.
 Further, changes in the decoherence parameter, $\tau$, only cause slight magnitude shifts in the flux and doesn't change the inference of the observations.
 
 In the cavity, coherent photons of energy $E_{ab}=E_a-E_b$  are generated each time the system relaxes from state $|a\rangle$ to $|b\rangle$.
 However, 
there is dissipation in the cavity mode due to stimulated emission. Since the cavity occupation, $n_l$, has to be kept constant,
this dissipation is proportional  to $\ln(\tilde n_l/n_l)$
\cite{goswami2013thermodynamics}. By taking care of this dissipation, the actual work done by the QHE can be written as,
\begin{eqnarray}
 W= E_{ab}-\alpha\ln\frac{\tilde{n}_l}{n_l},
\end{eqnarray}
with $\alpha$ as the proportionality factor. For a QHE with no driving $\alpha=k_BT_c$ \cite{goswami2013thermodynamics}.
 In this case, since there is a cyclic driving, the proportionality factor will depend on the driving protocol. However, per cycle, 
  the proportionality factor will be $\alpha=k_B/t_p\int_0^{t_p}T_c(t)dt$. 
  This is because, the cavity dynamics is equivalent for both the driven and non-driven cases with the only difference $T_c\rightarrow T_c(t)$. Using this definition, 
 we calculate the efficiency at maximum power (EMP) and focus on the expansion coefficients near equilibrium. 
 In order to calculate the EMP we maximize the power, $P=jW$ with respect to the energy $E_b$.
 Near equilibrium, we observe that the EMP goes linearly, $\eta_*=m\eta_c$. 
 It has been shown that $m=1/2$  and this value is universal in the close to equilibrium in the endoreversible regime\cite{PhysRevLett.102.130602,BroeckEMP}. 
 However, we see that $m=1/2$ only for $\phi$ being integer multiples of $\pi$ (i.e no PBp effects).
 For other phase differences $\phi>(<)\pi$, $m<(>)1/2$  as shown in Fig. (\ref{ph-linear}b). We further observe that the EMP is not restrictive 
  in the bounds $\eta_c/2\le \eta_*\le\eta_c/(2-\eta_c)$\cite{izumida2012efficiency,PhysRevE.87.012133} as seen from Fig. (\ref{ph-linear}b). When $\phi< (>)\pi$, the EMP
  crosses the upper (lower)  bound on $\eta_*$. 
  This can be explained from the oscillatory behavior of the geometric flux. Since $j_g>(<)0 \forall \phi<(>)\pi$, it adds (subtracts) to (from) the total flux, thereby 
   changing the position of $E_b$ where $P$ is maximum,  increasing (decreasing) the slope of $\eta_*$. 
 Deviations from universality in near equilibrium expression of $\eta_*$ has also been recently discussed in a classical heat engine with heat exchangers in the linear response regime\cite{PhysRevE.96.022119}.

\section{Steady State Fluctuation Theorem}
\label{s4}
In a general framework of open quantum systems, 
the FTs \cite{dhar2,uhrmp,campisi2011colloquium,utsumi2010bidirectional} have its manifestation in the generating function, $G(\lambda,t)$. 
   A symmetry of the type $G(\lambda,t)=G(-\lambda- F,t)$, popularly called the Gallavoti-Cohen (GC) symmetry \cite{Galla-Cohen,dhar2,uhrmp,max-uh} exists
   and is analogous to the FT. Any open quantum system with the GC symmetry satisfies this standard form of the SSFT \cite{uhrmp,max-uh}.
In open quantum systems, SSFT in the  long observation time reads,
$
 \ln\lim_{t\rightarrow\infty}[P(q,t)/P(-q,t)]=qF,
$
where $F$ is the thermodynamic 
 force that drives the system out of equilibrium and has recently been experimentally verified in quantum dots \cite{utsumi2010bidirectional}.
 The FT guarantees positive entropy production since the probability of observing negative flux is exponentially suppressed.
  The PDF is connected to the long time scaled generating function by the Gartner-Ellis-Varadhan theorem
 \cite{touchette2009large,varadhan1984large},
$
  P(q,t)\asymp \exp\{-t L(y)\}$ with $ 
  L(y)=\sup_\lambda^{}(y\lambda-S(\lambda)),$
 $L(y)$ being the large-deviation function (LDF), a Legendre transform of the scaled cumulant generating function with  $y=q/t$. The LDF obeys the symmetry,
$
 L(y)-L(-y)=-yF.
$ The symmetry in the LDF and the GC symmetry are implications of the SSFT\cite{uhrmp,dhar2}.
 
    \begin{figure}
\centering
\includegraphics[width=8.5cm]{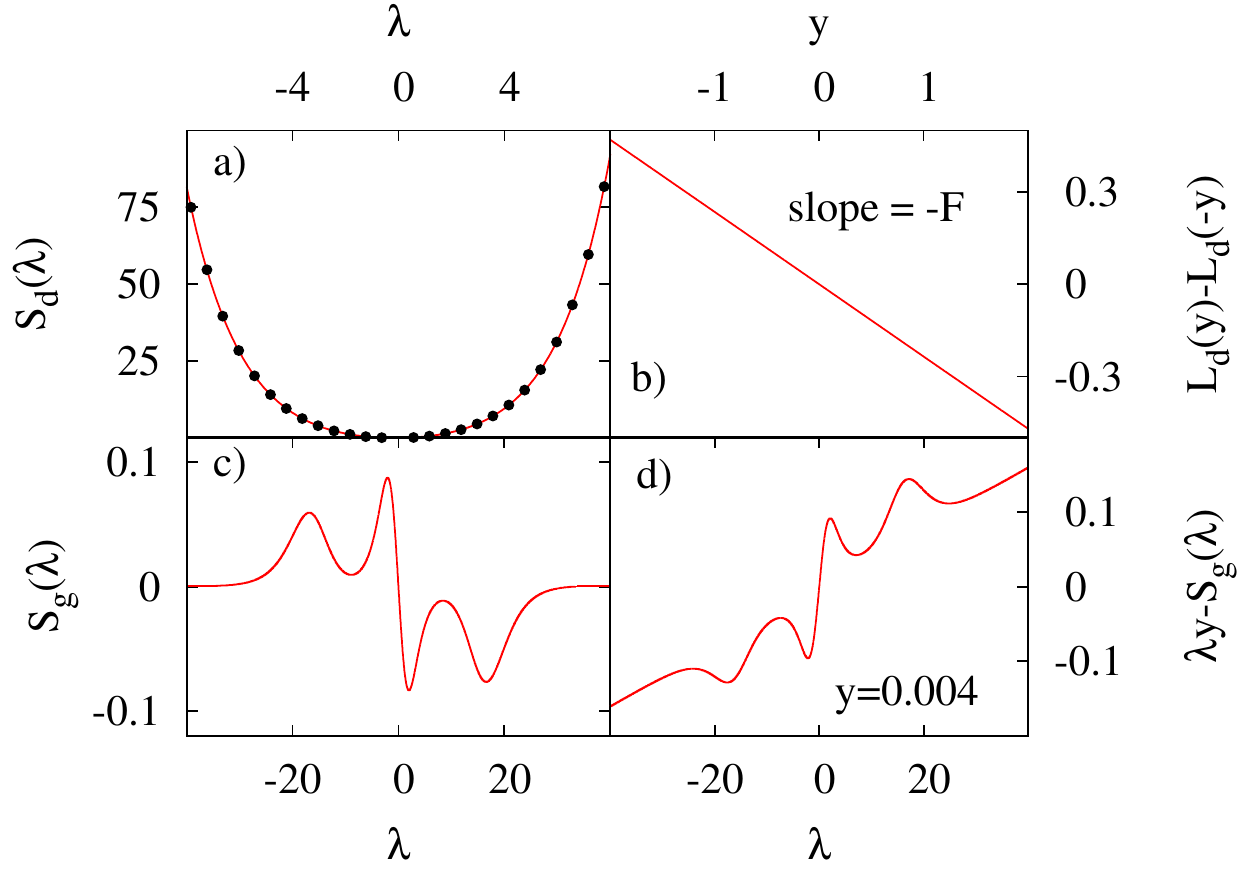}
\caption{a) GC symmetry  $S_d(\lambda) $(red)$=S_d(-\lambda-F)$ (dotted), $F=0.235$. b) Symmetry in the dynamic LDF with a slope of $-F$. c) Asymmetric behavior 
 of $S_g(\lambda)$ which  breaks SSFT. d) Non-convex  nature of the linearly shifted $S_g(\lambda)$.
 Parameters: $T_c=1,T_h=2.5,T_l=2,A=0.15,\omega=1.2, p_c=p_h=0,\phi=\pi/3,r=0.7,g=50,\tau=0.5$.}
\label{GCfig}
\end{figure}

In the QHE, we evaluate the GC symmetry by numerically calculating the 
 eigenvectors of the counting Liouvillian, Eq. (\ref{Liouvillian-lambda}). We find that the GC symmetry 
 holds only for the dynamic part, $S_d(\lambda)=S_d(-\lambda-F)$, Fig. (\ref{GCfig}a) as seen in some previous works
  with both heat and electron transport
 \cite{ren2010berry,hpg4} for any phase difference. Likewise, the symmetry in the LDF holds only for the dynamic part $L_d(y)-L_d(-y)=-yF$, Fig. (\ref{GCfig}b) . 
 The thermodynamic force in the QHE is (appendix),
 \begin{eqnarray}
 F=\displaystyle\ln \frac{\tilde n_l
 \int_0^{t_p}\tilde n_c n_hdt}{n_l\int_0^{t_p}n_c\tilde n_hdt}.
 \end{eqnarray}
 However, for a nonzero phase difference ($\phi\ne \pi,2\pi\ldots$),  $S_g(\lambda)$ also contributes to the PDF. 
Unlike the dynamic part, the GC symmetry doesn't hold for the geometric part, $S_g(\lambda)\ne S_g(-\lambda-R)$ (Fig. \ref{GCfig}c). Infact, there is no symmetry in $S_g(\lambda).$
This causes  the form of the SSFT to be violated, a direct consequence of finite PBp effects. 
Further, 
  $S_g(\lambda)$ as well as the linearly shifted function, $y\lambda - S_g(\lambda)$, (Fig. \ref{GCfig}d), are not  strict monotonously increasing functions. 
Hence, the standard Gartner-Ellis-Varadhan theorem cannot be used to evaluate the PDF anymore since a standard Legendre transformation cannot be performed. This inapplicability is due to the fact that the 
LDF is based on a saddle-point approximation valid for convex upwards or concave downward functions \cite{touchette2009large, varadhan1984large},
 which is  not the case here. 
Note that, the PDF  can be obtained by doing a standard inverse transform on the GF. 
We conclude that the standard form of the SSFT and the applicability of the large deviation theory is restricted to the case when 
PBp contributions are absent, i.e the curvature in Eq. (\ref{s-geo}) is zero.

\section{Conclusions}
\label{s5}
We investigated a periodically driven quantum heat engine by parametrically modulating the temperatures of the reservoir 
within an adiabatic quantum Markovian master equation approach. 
From a general framework of full counting statistics, we showed that the scaled cumulant generating function corresponding to the number of photons emitted in the cavity 
is separable into  dynamic and Pancharatnam-Berry phase-like 
contributions.  
The Pancharatnam-Berry phase-like contributions were shown to violate the steady state fluctuation theorem as well as the expansion coefficient and the  bounds on the universal expression
 for the efficiency at maximum power. We also showed that one cannot use 
  a large deviation technique to evaluate the probability distribution function. 
  These violations can however be recovered for a vanishing geometric curvature. Further we observed that,
  the effect of quantum coherences in optimizing the total flux can be nullified by having a dominant geometric contribution to the total flux.

\begin{acknowledgments}
   We acknowledge the support from the Max-Planck-Institute for the Physics of Complex Systems, Dresden, Germany and thank 
   Izaak Neri for going through the manuscript. 
   HPG would like to thank Prof. Udo Seifert for a helpful discussion.
 \end{acknowledgments}

\appendix*
\section{Derivation of QMME and PBp GF}

To second order
in coupling Hamiltonian, $\hat V$, 
the time evolution of
the reduced density matrix, $\rho(t)=\tr_B\tr_l\{\rho_T(t)\}$, is given by ($\hbar=1$),
\begin{align}
 \dot\rho(t)&=\displaystyle-i[\hat H_o,\rho(t)]-\int_0^t dt'\tr_B\tr_l[\tilde V(t),[\tilde V(t'),\rho_T(t')]],
\end{align}
with the interaction picture defined as $\tilde O(t)=e^{-i\hat Ht}\hat Oe^{i\hat Ht}$ and $\rho_T$ being the total density matrix. 
We assume separability of the $\rho_T(t)=\rho(t)\otimes\rho_B\otimes\rho_l$ where, 
 $ \rho_B,\rho_l$ are the  reservoir and cavity density matrices respectively. We further assume slow driving (adiabatic approximation) and that the bath correlations die fast (Markov approximations).
The timescales of the baths ($t_B$), system ($t$) and driving  ($t_d$) can be separated such that $t_{B}\gg t\gg t_d$. 
Under this adiabatic Markov approximation, using the definition of the coupling Hamiltonian and switching back to the Schrodinger picture, we get
\begin{widetext}
\begin{eqnarray}
 \dot\rho(t)&=&
 -i[\hat H_0,\rho(t)]-\pi\Omega_B\displaystyle\sum_{ij}\bigg[
 g_{ih}^{}g_{jh}^{*}\bigg\{\tilde n_h(\omega_{aj}^{},t)^{}\big{(}\hat B_{ja}\rho(t) \hat B_{ia}^\dag
 - \hat B_{ia}^\dag \hat B_{aj}^{}\rho(t)\big{)}\nonumber\\ &&+n_h(\omega_{aj}^{},t)^{}\big{(}
 \rho(t)\hat B_{ja}^{} \hat B_{ia}^\dag-\hat B_{ia}^\dag\rho(t) \hat B_{ja}
 \big{)}\bigg\}+
 g_{ih}^{*}g_{jh}^{}\bigg\{\tilde n_h(\omega_{aj}^{},t)^{}\big{(}\rho(t) \hat B_{ia}^\dag \hat B_{ja}
 - \hat B_{ia}^{} \rho(t)\hat B_{aj}^{\dag}\big{)}\nonumber\\ &&+n_h(\omega_{aj}^{},t)^{}\big{(}
 \hat B_{ia}^{} \hat B_{ja}^\dag\rho(t)-\hat B_{ja}^\dag\rho(t) \hat B_{ia}^{}
 \big{)}\bigg\}+
 g_{ic}^{*}g_{jc}^{}\bigg\{\tilde n_c(\omega_{bj}^{},t)^{}\big{(}\rho(t)\hat B_{bj}^{} \hat B_{bi}^\dag
 - \hat B_{bi}^\dag\rho(t) \hat B_{bj}^{}\big{)}\nonumber\\ &&+n_c(\omega_{bj}^{},t)^{}\big{(}
 \rho(t)\hat B_{bj}^{\dag} \hat B_{bi}^{}-\hat B_{bj}^{}\rho(t) \hat B_{bi}^\dag
 \big{)}\bigg\}+
 g_{ic}^{}g_{jc}^{*}\bigg\{\tilde n_c(\omega_{bj}^{},t)^{}\big{(}\hat B_{bi}^{} \hat B_{bj}^\dag\rho(t)
 - \hat B_{bj}^\dag\rho(t) \hat B_{bi}^{}\big{)}\nonumber\\ &&+n_c(\omega_{bj}^{},t)^{}\big{(}
 \rho(t)\hat B_{bj}^{\dag} \hat B_{bi}^{}-\hat B_{bi}^{}\rho(t) \hat B_{bj}^\dag
 \big{)}\bigg\}\bigg]\nonumber\\
&&
-\pi \Omega_l g^2\Big{[}\tilde{n}_l(\omega_{ab}^{})\big{[}\hat B^\dag_{ba}\hat B_{ba}\rho(t)
-2\hat B^{}_{ba}\rho(t)\hat B_{ba}^\dag
+\rho(t)\hat B^\dag_{ba}B_{ba}\big{]}\nonumber\\&&
-n_l(\omega_{ab}^{})\big{[}\hat B_{ba}\hat B^\dag_{ba}\rho(t)-2\hat B_{ba}^\dag\rho(t)\hat B^{}_{ba}
+\rho(t)\hat B_{ba}\hat B^\dag_{ba}\big{]}\Big{]}
\end{eqnarray}

\end{widetext}
Here, $\hat{H}_o$ is the bare Hamiltonian operator in the Hilbert space and $\omega_{ij}=E_i-E_j$.
The subscripts $h$ and $c$ represent the hot and cold thermal baths. 
$\Omega_k, k=B(l)$, the density of states for the baths (cavity) is assumed to be independent of frequency (wide-band approximation) and equal for both the hot $(h)$ and cold ($c$) 
thermal baths. The time-dependent Bose-Einstein functions for the baths are given by,
$n_x(\omega_{ij},t)=\tr_B\{\hat a_x^\dag\hat a_x\rho_x(t)\},x=h,c$  with $n_x(\omega_{ij},t)=(\exp(\omega_{ij}/k_BT_x(t))-1)^{-1}$ and 
$\tilde n_x=1+n_x$. We also have defined a  cavity occupation number as $n_l(\omega_{ab})=\tr_c\{\hat a_l^\dag\hat a_l\rho_l\}$=$
(\exp(\omega_{ab}/k_BT_l)-1)^{-1}$, where $\rho_l
$ is a fictitious cavity density matrix assumed to be held constant at a fictitious temperature $T_l$\cite{scully2003extracting,goswami2013thermodynamics}.
The density matrix elements are given by $\rho_{ij}=\langle i|\rho|j\rangle$. 
The density vector is given by
$|\rho\rangle=\{\rho_{11},\rho_{22},\rho_{aa},\rho_{bb},\Re(\rho_{12})\}$  
and contains both populations and the 
real part of coherences, i.e $\Re(\rho_{12})=\rho_{12}+\rho_{21}$. 
The coherence $\rho_{12}$ between states $|1\rangle$ and $|2\rangle $ is a thermally induced coherence arising because of the interactions with the hot and the cold baths. 
 The superoperator Liouvillian, $\breve{\mathcal{L}}(t)$ 
 is,
 \begin{widetext}
\begin{equation}
\breve{\mathcal{L}}(t)=\begin{pmatrix}
-\Gamma_{1c}^{}n_c^{}(t)-\Gamma_{1h}^{}n_h^{}(t)&0&\Gamma_{1h}^{}\tilde{n}_h(t)&\Gamma_{1c}^{}\tilde{n}_c(t)&-2\Gamma_{12}^{}(t)\\
0&-\Gamma_{2c}^{}n_c^{}(t)-\Gamma_{2h}^{}n_h^{}(t)&\Gamma_{2h}^{}\tilde{n}_h(t)&\Gamma_{2c}^{}\tilde{n}_c(t)&-2\Gamma_{12}^{}(t)\\
\Gamma_{1h}^{}n_h^{}(t)&\Gamma_{2h}^{}n_h^{}(t)&-\Gamma_h^{}\tilde n_h^{}(t)-g^2\tilde n_l^{}&g^2_{}n_l^{}&2\Gamma_{12h}^{}n_h(t)\\
\Gamma_{1c}^{}n_c(t)&\Gamma_{2c}^{}n_c(t)&g^2\tilde{n_l}&-g^2n_l-\Gamma_c^{}\tilde{n}_c(t)&2\Gamma_{12c}^{}n_c(t)\\
-\Gamma_{12}^{}(t)&-\Gamma_{12}^{}(t)&\Gamma_{12h}^{}\tilde{n}_h(t)&\Gamma_{12c}^{}\tilde{n}_c(t)&\bar g-\tau
\end{pmatrix}\\[2mm]
\end{equation}
\end{widetext}

where,
\clearpage
\begin{eqnarray}
 \Gamma_{12}(t)&=&\frac{1}{2}(\Gamma_{12c}n_c(t)+\Gamma_{12h}n_h(t))\\
 \bar g&=&-\frac{n_h(t)}{2}(\Gamma_{1h}+\Gamma_{2h})-
 \frac{n_c(t)}{2}(\Gamma_{1c}+\Gamma_{2c})\\
 \Gamma_{x}&=&\Gamma_{1x}+\Gamma_{2x}~~~x\in h,c
\end{eqnarray}
and,
\begin{eqnarray}
 \Gamma_{1x}^{}&=&\frac{\pi\Omega}{2}|g_{1x}^{}|^2~~~\Gamma_{12x}^{}=\frac{\pi\Omega}{2}|g_{1x}g_{2x}|^2~~~~x\in h,c
\end{eqnarray}
$\Gamma_{1x}(\Gamma_{2x})$ multiplied by the corresponding occupation factors represent 
  the rates of transition between $|1\rangle (|2\rangle)$ and $|a\rangle$ or $|b\rangle$. $\Gamma_{12x}$ is a measure of the  strength of
coherences. 
It is dependent on the relative orientation of the transition dipoles between an intermediate state $|a\rangle$ or $|b\rangle$
 and states $|1\rangle$ and $|2\rangle$. When the dipole vectors are perpendicular, the coupling vanishes and it is maximum
when dipoles are parallel. Thus $0\le \Gamma_{12x}\le\sqrt{\Gamma_{1x}\Gamma_{2x}}$. 
Accounting for these relative angles as controllable quantities,  two dimensionless parameters, $p_h$ and $p_c$ can be introduced such that
$0 \le p_h , p_c \le 1$, where subscripts h and c are used to keep track of contributions coming from couplings to the
hot and the cold baths, respectively. We can now rewrite $\Gamma_{12c} = rp_c ,\Gamma_{12h}= rp_h$ \cite{goswami2013thermodynamics,ScullyPNAS13092011}.

In the steadystate,  the photon fluctuations between system and cavity can be obtained using
the FCS method\cite{uhrmp} which involved the use of the characteristic or twisted generator
$\breve{\cal L}(\lambda,t)$. 
$\lambda$ is an auxiliary field which gets introduced as an exponential in 
the transitions involving photon exchange with the cavity.  In the QHE, the $\lambda$ dependence 
is carried by the time-independent matrix elements $\breve{\cal L}_{34}$ and $ \breve{\cal L}_{43}$, since these two elements are responsible for 
 photon exchange. Under the condition when 
 there is a positive flux of photons 
 into the cavity mode.
We now define a moment generating
 function, $G(\lambda,t)$ for the PDF corresponding to the net number of photons, $q$ exchanged
  between cavity and system.
 The equation of motion for $G(\lambda,t)$ is
\begin{equation}
  \label{M-app}
   \dot G(\lambda,t)=\langle \breve{\boldsymbol 1}| \breve{\cal L}(\lambda,t)|\rho(\lambda,t)\rangle.
  \end{equation} 
We have  denoted the time and counting-field dependent density vector as
$|\rho(\lambda,t)\rangle$ \cite{uhrmp} and 
\begin{widetext}
\begin{equation}
\breve{\mathcal{L}}(\lambda,t)=\begin{pmatrix}
-\Gamma_{1c}^{}n_c^{}(t)-\Gamma_{1h}^{}n_h^{}(t)&0&\Gamma_{1h}^{}\tilde{n}_h(t)&\Gamma_{1c}^{}\tilde{n}_c(t)&-2\Gamma_{12}^{}(t)\\
0&-\Gamma_{2c}^{}n_c^{}(t)-\Gamma_{2h}^{}n_h^{}(t)&\Gamma_{2h}^{}\tilde{n}_h(t)&\Gamma_{2c}^{}\tilde{n}_c(t)&-2\Gamma_{12}^{}(t)\\
\Gamma_{1h}^{}n_h^{}(t)&\Gamma_{2h}^{}n_h^{}(t)&-\Gamma_h^{}\tilde n_h^{}(t)-g^2\tilde n_l^{}&g^2_{}n_l^{}e^{-\lambda}&2\Gamma_{12h}^{}n_h(t)\\
\Gamma_{1c}^{}n_c(t)&\Gamma_{2c}^{}n_c(t)&g^2\tilde{n_l}e^{\lambda}&-g^2n_l-\Gamma_c^{}\tilde{n}_c(t)&2\Gamma_{12c}^{}n_c(t)\\
-\Gamma_{12}^{}(t)&-\Gamma_{12}^{}(t)&\Gamma_{12h}^{}\tilde{n}_h(t)&\Gamma_{12c}^{}\tilde{n}_c(t)&\bar g-\tau
\end{pmatrix}.\\[2mm]
\end{equation}
\end{widetext}
 For equal couplings, $\Gamma_{1h}=\Gamma_{2h}=\Gamma_{1c}=\Gamma_{2c}=r$, we recover Eq. (\ref{Liouvillian-lambda}). We now expand the density vector in the basis
 of the right eigenvector of $\breve{\cal L}(\lambda,t)$ with time dependent expansion coefficients $a_m(t)$\cite{ren2010berry,sin},
 \begin{equation}
 \label{rev basis}
 |\rho(\lambda,t\rangle=
 \displaystyle\sum_{m=1}^5a_m(t)e^{\int_0^t\zeta_m(\lambda,t')dt'} |R_m(\lambda,t)\rangle.                        
                          \end{equation}

Here $\zeta_m$ correspond to the five instantaneous eigenvalues of the characteristic Liouvillian.
Substituting Eq.(\ref{rev basis}) in Eq.(\ref{M-app}), and following the procedure outlined in the works\cite{ren2010berry,sin,hpg4},  we get,
 \begin{align}
G(\lambda,t)&=-\displaystyle\sum_{m=1}^5a_m(0)
  \langle {\bf 1}|R_m(\lambda,t)
 \rangle\nonumber\\
 &\times e^{
 {\int_0^{t}dt'[-\langle L_m(\lambda,t')|\dot R_m(\lambda,t')\rangle+\zeta_m(\lambda,t')]}}
 \end{align}
At long times, the contribution from
 all the more negative eigenvalues is  exponentially suppressed. Hence,
 at large times, 
 \begin{align}
 G(\lambda,t)&\approx a_o(0)\langle \breve{\boldsymbol 1}|R_o(\lambda,t)\rangle\nonumber\\
 &
 e^{\frac{t}{t_p}\int_0^{t_p}(\zeta_o(\lambda,t')
 -\langle L_o(\lambda,t'|\dot R_o(\lambda,t'\rangle)dt'},
\end{align}
 where we have used $t=nt_p$,  $n$ being the number of cycles and subscript $o$ corresponds to the dominating eigenvalue and eigenvectors. At the steady state, the scaled cumulant
  generating function is, 

  \begin{align}
  \label{cum}
   S(\lambda)&=\displaystyle\lim_{t \rightarrow\infty}\frac{1}{t}\ln G(\lambda,t)\\
 &= \lim_{n\rightarrow\infty}\frac{1}{n t_p}
 [\ln a_o(0)\langle I|R_o(\lambda,0)\rangle]\nonumber\\
 &+\frac{1}{t_p} \int_0^{t_p}\zeta_o(\lambda,t')dt'\nonumber\\
 &-\frac{1}{t_p}\int_0^{t_p}\langle L_o(\lambda,t')|\dot R_o(\lambda,t')\rangle dt'.
  \label{factored}
\end{align}
The first term in Eq. (\ref{factored}) is constant and goes to zero.
The scaled cumulant generating function can now be expressed as a sum of
a dynamic ($S_d(\lambda)$)and geometric ($S_g(\lambda)$) scaled cumulant generating functions given by Eqs. (\ref{s-dyn}) and (\ref{s-geo1}).

For $\phi=0,n\pi,n\in$ integers, Eq. (\ref{s-geo}) is zero. In this limit $j_g=0$ and 
the SS dynamic flux  can also be obtained using $j_d=\langle \breve{\boldsymbol 1}|{\breve{\cal L}}_l(t)|\rho_s\rangle$ 
 where ${\breve{\cal L}}_l(t)$ is the Liouvillian containing elements pertaining to cavity alone, i.e 
  the matrix elements $\breve{\cal L}_{33}, \breve{\cal L}_{34}, \breve{\cal L}_{43}$ and $ \breve{\cal L}_{44}$. $|\rho_s\rangle$ is the steadystate density vector that can be obtained
  by solving
 $\breve{\cal L}|\rho(t)\rangle=0$. Substituting the steady state values of the populations and coherence in the expression for 
 $j_d$, and integrating $j$ over a cycle, one  can get an analytical expression for the dynamic flux per cycle. From this expression, 
 the thermodynamic force $F$ can be identified. Note that although $F$ can be analytically obtained only for $\phi=0,\pi,2\pi...$, 
 the symmetry $S_d(\lambda)=S_d(-\lambda-F)$ holds for 
 all values of $\phi$.

\bibliography{references.bib}

  \end{document}